\documentclass[10pt]{iopart}

\usepackage{amsfonts}
\usepackage{amssymb}
\usepackage{xfrac}
\usepackage{graphicx}
\usepackage{booktabs,array}
\usepackage{comment}
\usepackage[skip=2pt,font=footnotesize]{caption}
\usepackage{titlesec}
\usepackage[flushleft]{threeparttable}
\usepackage{color}
\usepackage[normalem]{ulem}
\usepackage{tablefootnote}

\newcolumntype{L}[1]{>{\raggedright\arraybackslash}m{#1}}
\newcolumntype{C}[1]{>{\centering\arraybackslash}m{#1}}
\newcolumntype{R}[1]{>{\raggedleft\arraybackslash}m{#1}}

\titleformat*{\section}{\normalsize\bfseries\rmfamily}
\titleformat*{\subsection}{\normalsize\bfseries\rmfamily}
\titleformat*{\subsubsection}{\normalsize\bfseries\rmfamily}

\graphicspath{{./Figures/},{./Workflows/}}
\DeclareGraphicsExtensions{.eps,.pdf,.jpg,.png}

\usepackage{subfig}
\usepackage{verbatim}
\usepackage{caption}
\usepackage{dcolumn}
\usepackage{bm}

\usepackage[backend=bibtex,maxcitenames=10,sorting=none,style=numeric,natbib=true]{biblatex}
\begin{document}

\title{Multiple-isotope pellet cycles captured by turbulent transport modelling in the JET tokamak}

\author[M. Marin, J. Citrin, L. Garzotti, M. Valovic, C. Bourdelle, Y. Camenen, F. J. Casson, A. Ho, F. Koechl, M. Maslov and JET Contributors]{M. Marin$^1$, J. Citrin$^1$, L. Garzotti$^2$, M. Valovi\v{c}$^2$, C. Bourdelle$^3$, Y. Camenen$^4$, F. J. Casson$^2$, A. Ho$^1$, F. Koechl$^2$, M. Maslov$^2$ and JET contributors$^*$}

\address{$^1$ DIFFER - Dutch Institute for Fundamental Energy Research, Eindhoven, the Netherlands}
\address{$^2$ CCFE, Culham Science Centre, Abingdon, OX14 3DB, United Kingdom of Great Britain and Northern Ireland}
\address{$^3$ CEA, IRFM, F-13128 Saint-Paul-lez-Durance, France}
\address{$^4$ CNRS, Aix-Marseille Univ., PIIM UMRT7345, 13397 Marseille Cedex 20, France}
\address{$^*$ See the author list of 'Overview of the JET preparation for Deuterium-Tritium Operation' by E. Joffrin et al 2019 Nucl. Fusion 59 112021 https://iopscience.iop.org/article/10.1088/1741-4326/ab2276 }

\date{\today}

\begin{abstract}

For the first time the pellet cycle of a multiple-isotope plasma is successfully reproduced with reduced turbulent transport modelling, within an integrated simulation framework. Future nuclear fusion reactors are likely to be fuelled by cryogenic pellet injection, due to higher penetration and faster response times. Accurate pellet cycle modelling is crucial to assess fuelling efficiency and burn control. In recent JET tokamak experiments, deuterium pellets with reactor-relevant deposition characteristics were injected into a pure hydrogen plasma. Measurements of the isotope ratio profile inferred a Deuterium penetration time comparable to the energy confinement time. The modelling successfully reproduces the plasma thermodynamic profiles and the fast deuterium penetration timescale. The predictions of the reduced turbulence model QuaLiKiz in the presence of a negative density gradient following pellet deposition are compared with GENE linear and nonlinear higher fidelity modelling. The results are encouraging with regard to reactor fuelling capability and burn control.

\end{abstract}

\maketitle
\ioptwocol

\section{Introduction} \label{Introduction}

In present tokamaks, particle fuelling is mainly provided by neutral gas puffing from the plasma periphery and from Neutral Beam Injection (NBI). Gas fuelling may be rendered ineffective in future reactors due to increased neutral opacity, while the particle source from the NBI will be relatively small. A viable alternative as a primary fuelling technique is the injection of cryogenic pellets \cite{Doyle2007}, with higher penetration and faster response times. Pellet mass, injection speed and frequency can be jointly adjusted to optimize the particle source and provide fuelling in the plasma core, where the pellet is ablated.
In ITER, for example, pellets of mass between $ 2 $ and $ 5.5 \cdot 10^{21} $ atoms with frequency between $ 1.5 $ and $ 3.5 $ Hz respectively should be sufficient to maintain the density required for a Q=10 baseline ELMy H-mode scenario at $ 15 $ MA.

Active research on pellet fuelling focuses on its compatibility with integrated plasma scenario constraints, including control of MagnetoHydroDynamic (MHD) modes such as Edge Localised Modes (ELMs), plasma exhaust, core turbulent transport, and desired isotope composition. Previous integrated tokamak plasma simulation (integrated modelling) including pellets focused on various aspects of the pellet cycle: improved confinement regimes \cite{Frigione2007, Garzotti2006a, Romanelli2004}, edge and fuelling requirements \cite{Polevoi2018}, the impact of fuelling on divertor heat-loads \cite{Wiesen2017, Garzotti2019} and the extrapolation of pellet penetration and transport \cite{Parail2009}.

Pellet fuelling and simultaneous ELMs mitigation have been demonstrated experimentally \cite{Valovic2018, Valovic2020, Lang2012}, ensuring the viability of this fuelling method. Regarding turbulent transport, the pellets have a significant impact. During the ablation phase of the pellet cycle, the density and temperature profiles are transiently modified, changing the micro-instability properties of the discharge. While the heightened negative radial density gradient that develops in the region outside the pellet ablation region is expected to destabilize Trapped Electron Modes (TEM) and lead to a strong outward particle flux \cite{Garzotti2003}, the positive density gradient that develops at radii within the ablation location may stabilize Ion Temperature Gradient (ITG) driven turbulence. This was observed for example in the Mega Amp Spherical Tokamak (MAST) \cite{Garzotti2014}. The stabilization was instead counteracted by a larger $ R/L_{T} $ again in MAST, with different plasma conditions, \cite{Valovic2008} and in a similar Joint European Torus (JET) experiment \cite{Tegnered2017}. $ R $ here is the major radius, while $ L_T $ is the logarithmic temperature gradient $ L_T = [ \partial(ln T)/ \partial r]^{-1} $.

In reactors, pellet injection with varying isotope ratios will be used to maintain the desired concentrations of deuterium and tritium in the core; equal ratios ensures maximal fusion power, and burn control is achieved by modifying the relative isotope concentrations. Understanding the timescales for the transport of different isotopes following modification of the pellet isotope composition is fundamental for understanding and predicting burn control. Since the electron and ion particle fluxes must always be equal (ambipolarity), differences in their transport can only be observed experimentally in presence of multiple ion types, e.g. hydrogenic isotopes. Previous experiments observed a fast mixing of T-trace in the Tokamak Fusion Test Reactor (TFTR) \cite{Efthimion1995}, while modelling found helium diffusion coefficients on the order of the effective heat conductivity in ITER simulations \cite{Angioni2009a}. Theoretical analysis recently explained the fast isotope mixing by $D_i/D_e>1$ and $|V_i|>|V_e|$ in ITG dominated plasmas \cite{Bourdelle2018a}, where $ D_{s} $ and $ V_{s} $ are the species dependent diffusion and pinch coefficients respectively. In a multi-ion plasma the different ions can interchange at different timescales to the electron particle transport. The opposite relation holds for TEM dominated regimes, as shown experimentally in the Large Helical Device (LHD) \cite{Ida2020}.

Previous multiple-isotope experiments at JET allowed a detailed investigation of ion particle transport \cite{Maslov2018}, suggesting fast isotope mixing. Those experimental observations were successfully reproduced in stationary-state, multiple-isotope integrated modelling \cite{Marin2020}, applying the quasilinear gyrokinetic transport model QuaLiKiz \cite{Bourdelle2016, Citrin2017}, strengthening QuaLiKiz validation in multiple-isotope regimes.

The fast mixing will be most prevalent during transient states, such as during pellet injections, due to the significant modifications of the local density gradients created by the short ablation time of the pellets. Modifying the pellet isotope ratio compared to the background isotope ratio rapidly changes the core isotope mix without affecting the time averaged electron profile. An experiment was performed at JET precisely with the aim of using pure deuterium pellets to control the core isotope ratio, starting from a pure hydrogen plasma \cite{Valovic2019}, and is investigated next.

Relevant parameters of the discharge under investigation are shown in table \ref{tab:dischargedetails}. In this experiment the size of the pellets, scaled to the plasma volume, lead to shallow deposition and transient inverted density profile, similarly to what is expected in ITER. $ \sim 10 \% $ of the pellet was ablated in the pedestal region, between $ 0.95 < \rho < 1.0 $, where the ad-hoc pedestal model is used.  Most of it was instead ablated inside the pedestal top, where the transport is predicted by QuaLiKiz, with $ \sim 88 \% $ between $ 0.6 < \rho < 0.95 $". $ \rho $ here indicates the normalised toroidal flux coordinate $ \rho_{tor} = (\frac{\psi_{tor}}{\psi_{tor, LCFS}})^{\frac{1}{2}}. $

\begin{table*}
    \caption{Key parameters of JET shot $ \# 91393 $. $ I_{p} $ and $ B $ are the total plasma current and the on-axis magnetic field, $ P $ and $ \Phi_{pel} $ represent injected heating power and pellet particle fuelling rate. $ \beta_N $ is normalized plasma beta defined as $ 2 \mu_{0} \frac{2}{3} a_{min} * W_{tot} / (I_{p} * V * B_{geo}) $, with $a_{min} = 0.5 * (R_{out,LCFS} - R_{in,LCFS})$, $ B_{geo} $ the vacuum toroidal field at the geometric plasma centre and $ V $ the volume}
    \centering
    \smallskip
    \begin{tabular}{c|c|c|c|c|c|c|c|c}
	    $ I_{p} $ [MA] & B [T] & $ Z_{eff} $ & $ P_{NBI} $ [MW] & $ P_{ICRH} $ [MW] & $ \Phi_{H2,gas} [10^{21} at/s] $ & $ f_{pel} [Hz] $ & $ \Phi_{pel} [10^{21} at/s] $ & $ \beta_N $ \\
        \hline
	    1.4 & 1.7 & 1.4 & 8.4 & 3 & 6.7 & 11.4 & 8.2 & ~1.1 \%\\

	\end{tabular}
	\label{tab:dischargedetails}
\end{table*}

The experiment managed to reach the desired core isotope composition, measured by Balmer-alpha Charge Exchange (CX) spectroscopy and D-D neutron rate. A rapid increase in the neutron rate following the initial pellet injection was observed. The delay between the start of the pellet ablation and the local peak in the neutron rate was $ \sim 100 ms $, which is comparable with the energy confinement time, $ \sim 120 ms $ for this experiment. This timescale is much faster than the particle confinement time, which is $ \sim 600 ms $, and indicated fast isotope mixing. In this interpretive analysis, the isotope particle transport coefficients were determined by interpretative modelling, using the semi-empirical Bohm/Gyrobohm turbulent transport model and matching the transient response of the thermal D-D neutron rates \cite{Valovic2018}. 

The key observation was that $D_{D}/\chi_{eff} \sim 1$ was inferred at the beginning of the pellet train, where $ D_{D} $ is the diffusion coefficient for Deuterium and $ \chi_{eff} $ is the effective heat conductivity. Since $D_{e}/\chi_{eff} \ll 1$ is expected in the experiment, this finding implies a large $D_{D}/D_{e}$, indeed consistent with the fast isotope mixing.

This paper demonstrates, for the first time, that multiple pellet cycles and the associated fast deuterium penetration can be captured by turbulent transport models within an integrated modelling framework, in an ITER-relevant pellet deposition regime.

\section{Integrated modelling} \label{Methods}

\begin{figure}

  \subfloat[]{\includegraphics[width=.57\linewidth]{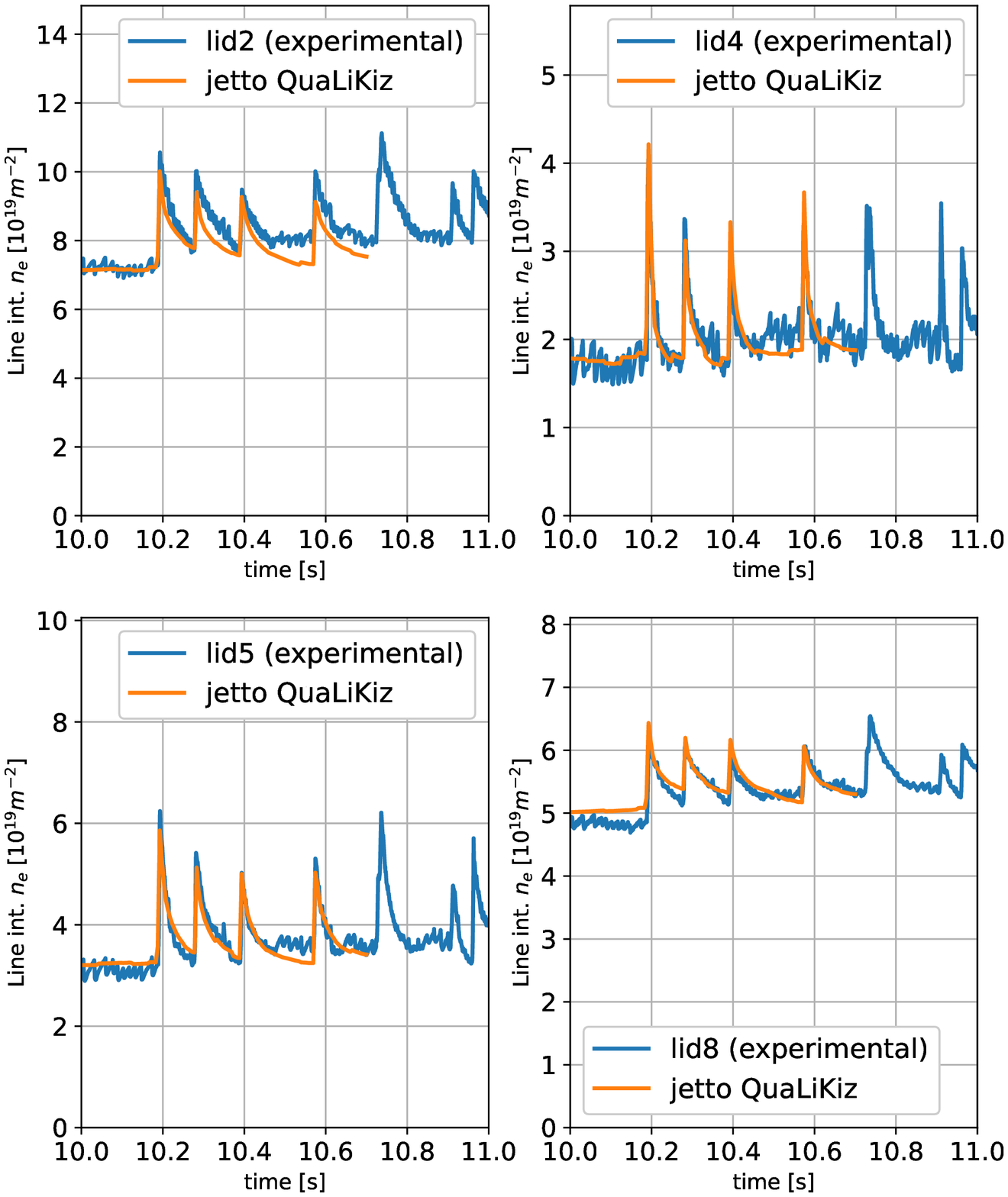}}
  \subfloat[]{\includegraphics[width=.44\linewidth]{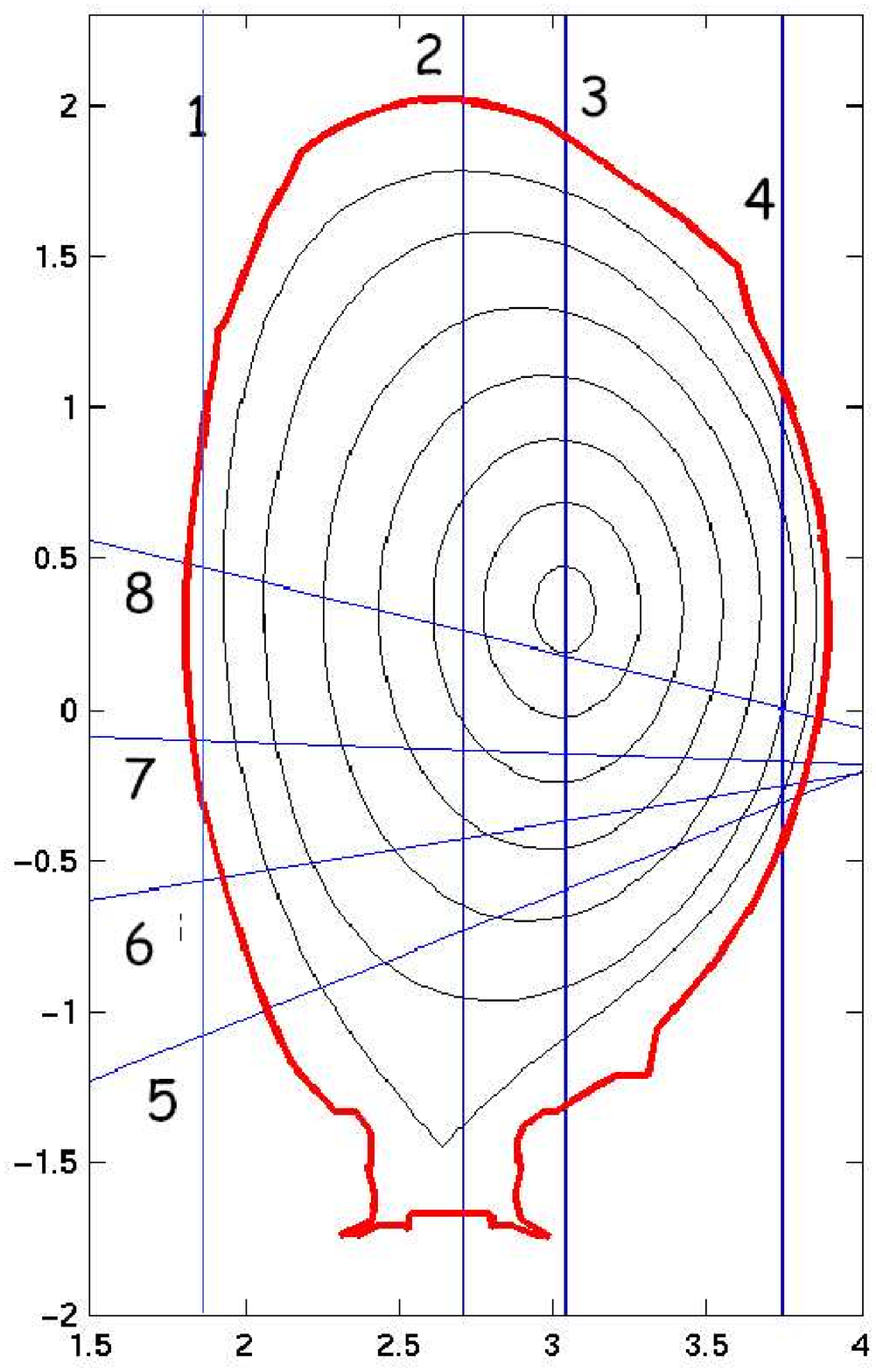}}

  \caption{a) Four different experimental interferometer lines (solid blue lines) for shot $\# 91393 $, compared with a synthetic diagnostic in JINTRAC (solid orange lines). The pellets are injected at $ t = 10.187, 10.278, 10.390, 10.572 $. b) Sketch showing the position of the lines of sight of the interferometer at JET}
  \label{fig:interferometer_lines}
\end{figure}

The modelling is performed within the JINTRAC \cite{Romanelli2014} framework, using JETTO as the 1.5D core transport solver. NCLASS \cite{Houlberg1997} is used as the neoclassical transport model and QuaLiKiz as the turbulent transport model. The initial electron density and ion and electron temperature profiles are obtained through Gaussian Process Regression (GPR) \cite{ho2019} on the experimental data, averaged for 200ms immediately before the first pellet. PENCIL \cite{Challis1989} and PION \cite{Eriksson1993} are used for NBI and ICRH heating respectively, FRANTIC \cite{Tamor} for the neutral source and HPI2 \cite{Koechl2012} for the pellet ablation. The impurities are modelled with SANCO \cite{Alper} and the magnetic equilibrium is evolved self-consistently using ESCO \cite{Jetto_manual}. HPI2 \cite{Koechl2012} is used as the pellet deposition model.

EFIT++ is used to obtain the last-closed-flux-surface boundary conditions for ESCO. The initial current profile is obtained by relaxing an initial EFIT \cite{Search1985} reconstruction until the safety factor $(q) = 1$ surface approaches the observed sawteeth inversion radius. This current profile evolution is carried out using ESCO for the magnetic equilibrium and NCLASS for the resistivity, while keeping density and temperature profiles fixed in time.

Beryllium and Nickel, consistently observed in JET discharges with Ion Cyclotron Resonance Heating (ICRH) \cite{Czarnecka2011, Sertoli2019}, are chosen as impurities to match both dilution and $ Z_{eff} $. Given that the radiated power in the core is below $ 20 \% $ of the heating power, the Tungsten content is inferred to be low. In order to reduce computational expense it is not included the simulation, as done in previous works \cite{Breton2017, Casson2020}.

Since QuaLiKiz is restricted to electrostatic turbulence, an ad-hoc model is employed to simulate the level of electromagnetic (EM) stabilization, as done previously \cite{Casson2020}. The ion temperature gradient passed to QuaLiKiz is multiplied by the local value of $ \frac{W_{thermal}}{W_{thermal}+W_{fast}} $, based on the expected correlation between fast ion content and EM-stabilization of ITG turbulence in NBI and ICRH heated plasmas. Here $W_{thermal}$ and $ W_{fast} $ are the contributions to the total energy content from the thermal and the fast particles respectively. Dedicated linear runs with the gyrokinetic code GENE \cite{Jenko2000} suggested a significant impact of EM-stabilisation on the linear growth rates at inner radii, justifying the inclusion of this effect.

Electron Temperature Gradient (ETG) driven modes are included in the simulation. The ETG transport levels are tuned to a a single-scale GENE nonlinear run and a simple multi-scale rule is used. The ETG heat flux is multiplied by 

\begin{equation}
f_{multi-scale} = \frac{1}{1+exp[-\frac{1}{5}({\frac{\gamma_{ETG-max}}{\gamma_{ITG-max}}-\sqrt{\frac{m_{i}(1)}{m_{e}}}})]} 
\end{equation}

with $ \gamma_{ETG-max} $ and $ \gamma_{ITG-max} $ being the maximum growthrates for ETG and ITG respectively, while $ m_{e} $ and $ m_{i}(1) $ are the electron mass and the mass of the first ion in QuaLiKiz. This ensures that non-negligible ETG fluxes arise only when the ETG growth rates are at least a mass ratio larger than their ITG counterparts, a rule-of-thumb derived from nonlinear multiscale simulations \cite{Howard2016}.

The radial zone incorporating QuaLiKiz-predicted turbulent transport is $ 0.2 < \rho < 0.95 $ For $ \rho < 0.2 $ modest heat and particle ad-hoc transport is artificially added. This term takes into account the average transport originating from intermittent (1,1) MHD activity (sawteeth). The pedestal region, $ 0.95 < \rho < 1 $, is out of the scope of the QuaLiKiz model, due to the nature of the pedestal turbulence and its suppression, as well as intermittent MHD activity (ELMs).

\begin{figure}
	\centering
	\includegraphics[width=1.0\linewidth]{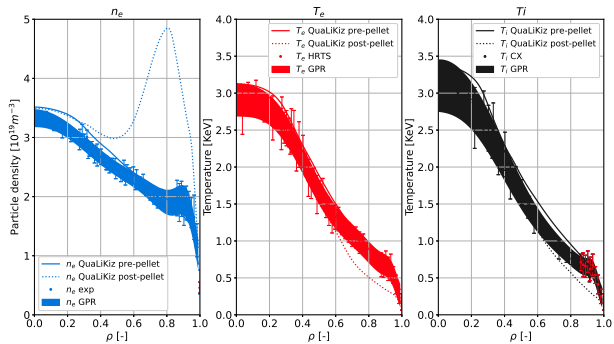}
	\caption{The shaded area represents the GPR confidence interval, with the experimental data averaged between $ 9.5 s < t < 10.15 s $. The solid line is the JINTRAC-QuaLiKiz prediction for density and temperature profiles before the first pellet ($t = 10.18 s$), after $ \sim 2 $ particle confinement times of relaxation. The boundary conditions at the Last Close Flux Surface (LCFS) are: $ n_{e_{1}} =0.7\cdot 10^{19} [m^{-3}]$, $ T_{e} = T_{i} = 100 [eV] $ The dotted lines show the profiles just after the first pellet injection ($t = 10.19 s$)}
	\label{fig:eps2019profiles}
\end{figure}

The perturbation caused by the pellet modifies the profiles in the pedestal region in a non trivial way. The pedestal is therefore evolved using a "continuous ELM model". The idea here is simply to match the temperature and density evolution at the top of the pedestal and provide appropriate core boundary conditions. The transport in the Edge Transport Barrier (ETB) is treated by the continuous ELM model described in \cite{Parail2009}, which mimics the limiting effect of the ELMs on the pressure gradient in the ETB by introducing additional transport averaged over time and clamps the normalized pressure gradient in the ETB, $ \alpha $, at a prescribed critical value, $ \alpha_{c} $, fitted to the experimental value. The parameters in this model are adjusted to match the interferometer measurement of the line of sight looking at the pedestal, indicated with '4' in Figure \ref{fig:interferometer_lines} (b), with a synthetic diagnostic within JINTRAC. The result is seen in Figure \ref{fig:interferometer_lines} (a). This decision resulted in a $ n_{e} $ at the top of the pedestal on the lower end of the errorbar with respect to the GPR fit, as shown in Figure \ref{fig:eps2019profiles}, which suffers from lower precision in that region due to the presence of ELMs. These parameters are kept constant during the simulation.

Outward particle convection is added as $ v=v_{0} \times exp\{-(t-t_{pel})/\tau+(r/a-1)/\Delta\} $ where $ v_{0} $, $ \tau $ and $ \Delta $ are parameters fitted to match the final total density. The need for this term, which mimics the extra ELMs density pump out in the presence of pellets, was recognized in previous works \cite{FKoechlIAEA2018}. $ \tau $ ensures that this term is only non negligible for a short period after the pellets injection, around 15ms in this case, while $ \Delta $ limits the effect to the radial zone close to the pedestal. Therefore, this term mostly overlaps with the continuous ELM model, with little effect on the QuaLiKiz predictions. The evolution of the density is still largely controlled by the evolution of the pedestal top, with this term ending up only slightly modifying the final density peaking.

Finally, it is worth noting that the pedestal stability to peeling - ballooning modes is modified by the pellets. The continuous ELM model keeps the pressure constant and since the density evolution at the top of the pedestal is ultimately the result of a fit, there is a difference in the temperature. The experimental electron temperature rises slightly slower in the experiment than in the simulation, but this difference never exceeds $ \sim 50eV $, so it is not expected to have a large impact on the profiles.

The pellet cycle modelling initial condition correspond to the stationary state JINTRAC-QuaLiKiz solution of the experimental configuration, after relaxing for $ \sim 2 $ particle confinement times, just before the beginning of the pellet train. This is shown in Figure \ref{fig:eps2019profiles}.

\section{Comparison with the experiment} \label{Comparison}

The good agreement shown in Figure \ref{fig:eps2019profiles} is reached in the electron temperature $ T_{e} $ and the ion temperature $ T_{i} $. The peaking of the electron density $ n_{e} $ is slightly overestimated. This general agreement provides confidence that the turbulent regime is correctly captured. The slight trend for improved predicted core confinement for this hydrogen plasma, compared to the measured profiles, may be a result of QuaLiKiz gyroBohm scaling. The nonlinear saturation rule was fit to deuterium plasma gyrokinetic simulations, while observations and nonlinear gyrokinetic simulations show an inverse isotope confinement scaling, with worse confinement for hydrogen \cite{Maggi2018}.

Four pellets are modelled, from $ t = 10.0 s $ to $ t = 10.7 s $. The model proved robust in responding to the significant changes in the profiles introduced by the pellets. All the measured interferometer lines of sight are compared with a synthetic diagnostic, resulting in general good agreement as shown in Figure \ref{fig:interferometer_lines} (a). The gradients for two radial points before and after the first pellet are listed in table \ref{tab:gradientsprepost} and shown in the dotted lines of Figure \ref{fig:eps2019profiles}. A direct comparison between the experimental and modelled neutron rate, which is a direct marker of inner-core deuterium content, is carried out. 

\begin{figure}
	\centering
	\includegraphics[width=1.0\linewidth]{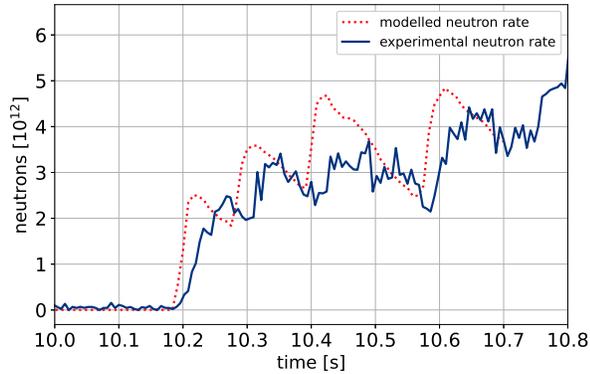}
	\caption{Measured neutron rate (blue solid line) vs the simulated neutron rate (red dashed line). The fast timescale of D penetration is captured by the modelling}
    \label{fig:neutron_rate}
\end{figure}

\begin{figure}
	\centering
	\includegraphics[width=1.0\linewidth]{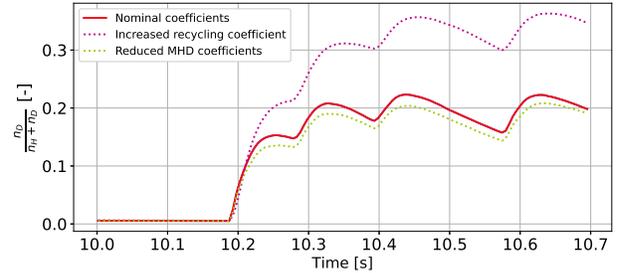}
	\caption{The red solid line is the $ n_{D}/n_{e} $ ratio at $ \rho = 0.15 $ for JET shot $ \# 91393 $ as predicted by JINTRAC-QuaLiKiz with nominal pedestal coefficients. The magenta dotted line has increased deuterium recycling coefficient in FRANTIC and the green dotted line has reduced scale factors in the continuous ELM model. The fast penetration of D is resilient to the precise tuning of the edge models}
	\label{fig:nDevolutioncentre}
\end{figure}

\begin{table*}
    \caption{Density and temperature gradients before ($ t=10.185 s $) and after ($ t=10.189 s $) the first pellet. $ 1/L_p $ is defined as $- \frac{1}{p} \, \frac{dp}{dr} $, with r being the minor radius and p a generic quantity. The two radial positions were chosen to isolate the large positive and negative density gradients induced promptly after a pellet deposition}
    \centering
    \smallskip
    \begin{tabular}{c|ccc|ccc}
	    & \multicolumn{3}{c|}{Pre-Pellet} & \multicolumn{3}{c}{Post-Pellet} \\ 

	    Gradient & $ R/L_{T_{i}} $ & $ R/L_{T_{e}} $ & $ R/L_{n_{e}} $ & $ R/L_{T_{i}} $ & $ R/L_{T_{e}} $ & $ R/L_{n_{e}} $ \\ 
        \hline
		$ \rho = 0.68 $ & 7.4 & 7.7 & 2.8 & 14.4 & 18.1 & -11.4 \\ 

		$ \rho = 0.85 $ & 11.1 & 12.2 & 5.6 & 9.5 & 8.8 & 14.4 \\ 

	\end{tabular}
	\label{tab:gradientsprepost}
\end{table*}

\begin{figure}
	\centering
	\includegraphics[width=1.0\linewidth]{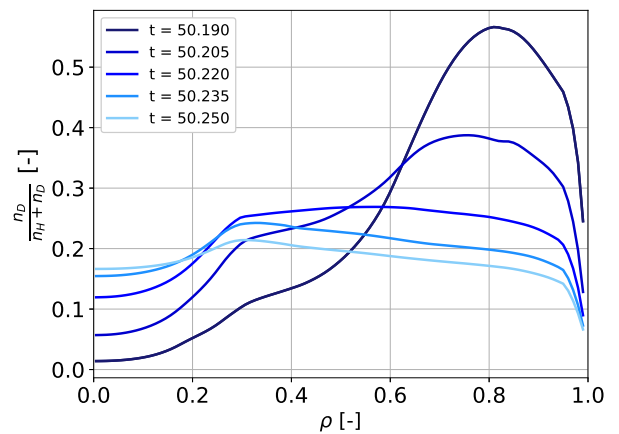}
	\caption{Evolution of the $ \frac{n_D}{n_e} $ profile after the first pellet injection. The profile is plotted every 15ms, with lighter shades of blu corresponding to increasing time}
	\label{fig:nDevolution}
\end{figure}

The $ \frac{n_{D}}{n_{e}} $ ratio is heavily dependent on the edge transport conditions, which are not predicted in the simulations. The edge transport model free parameters are adjusted to match the final neutron rate to the experimental value. These parameters are the constant recycling coefficient for deuterium in FRANTIC and the minimum deuterium transport coefficient in the ETB model. Both have a very similar effect on deuterium concentration and are here used as knobs, so the precise values should not be expected to hold physical significance. The deuterium content at the LCFS is increased linearly starting from the first pellet, reaching the experimentally measured $ 20\% $ by the end of the simulation. The experimental and modelled neutron rate are found to be in good agreement, as can be seen in Figure \ref{fig:neutron_rate}. The $ \frac{n_D}{n_e} $ evolution after the first pellet is shown in figure \ref{fig:nDevolutioncentre}, figure \ref{fig:nDevolution}. As expected, the ratio quickly relaxes to a rather flat profile.

\begin{figure*}

  \subfloat[]{\includegraphics[width=.33\linewidth]{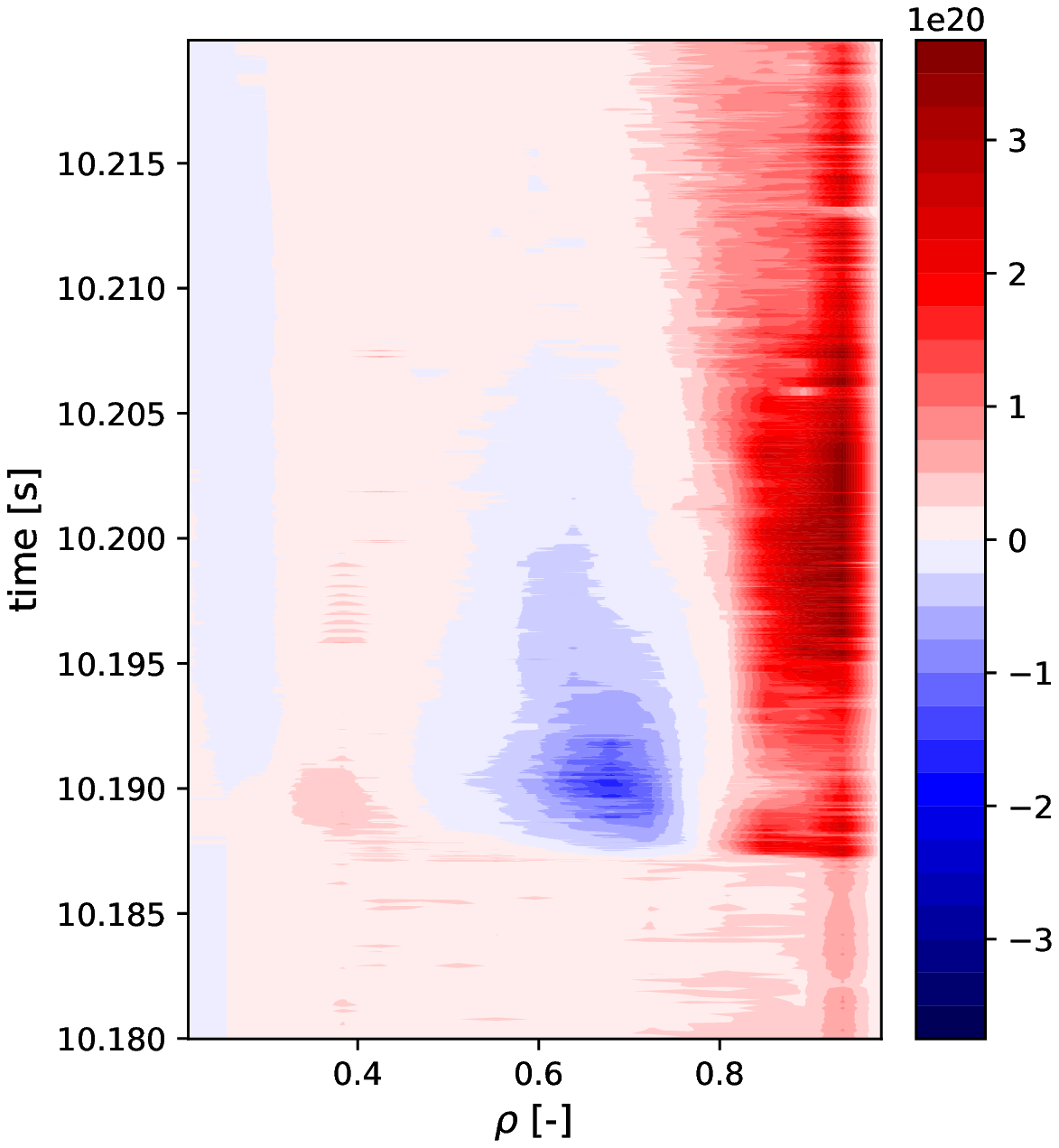}}
  \subfloat[]{\includegraphics[width=.33\linewidth]{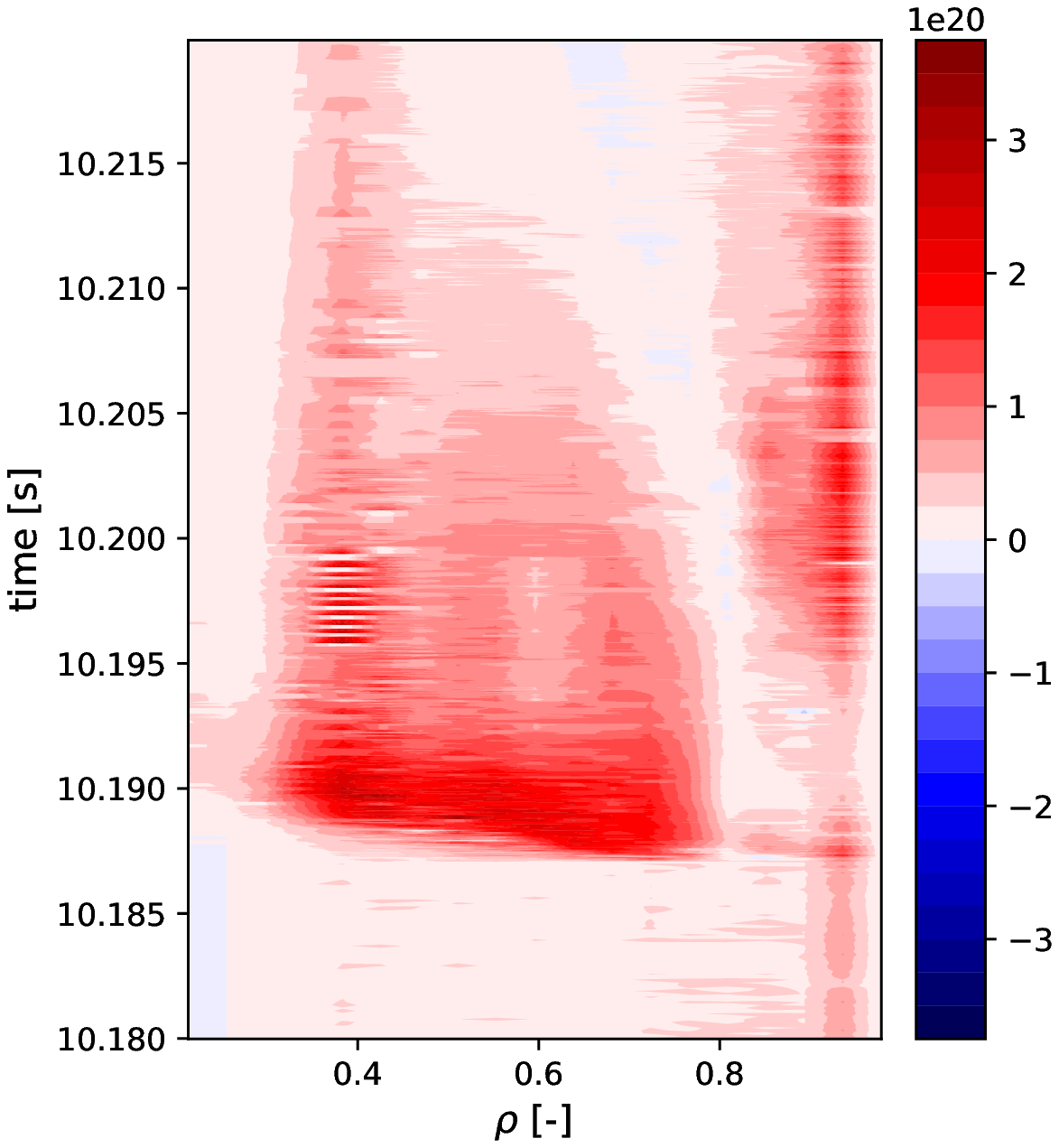}}
  \subfloat[]{\includegraphics[width=.33\linewidth]{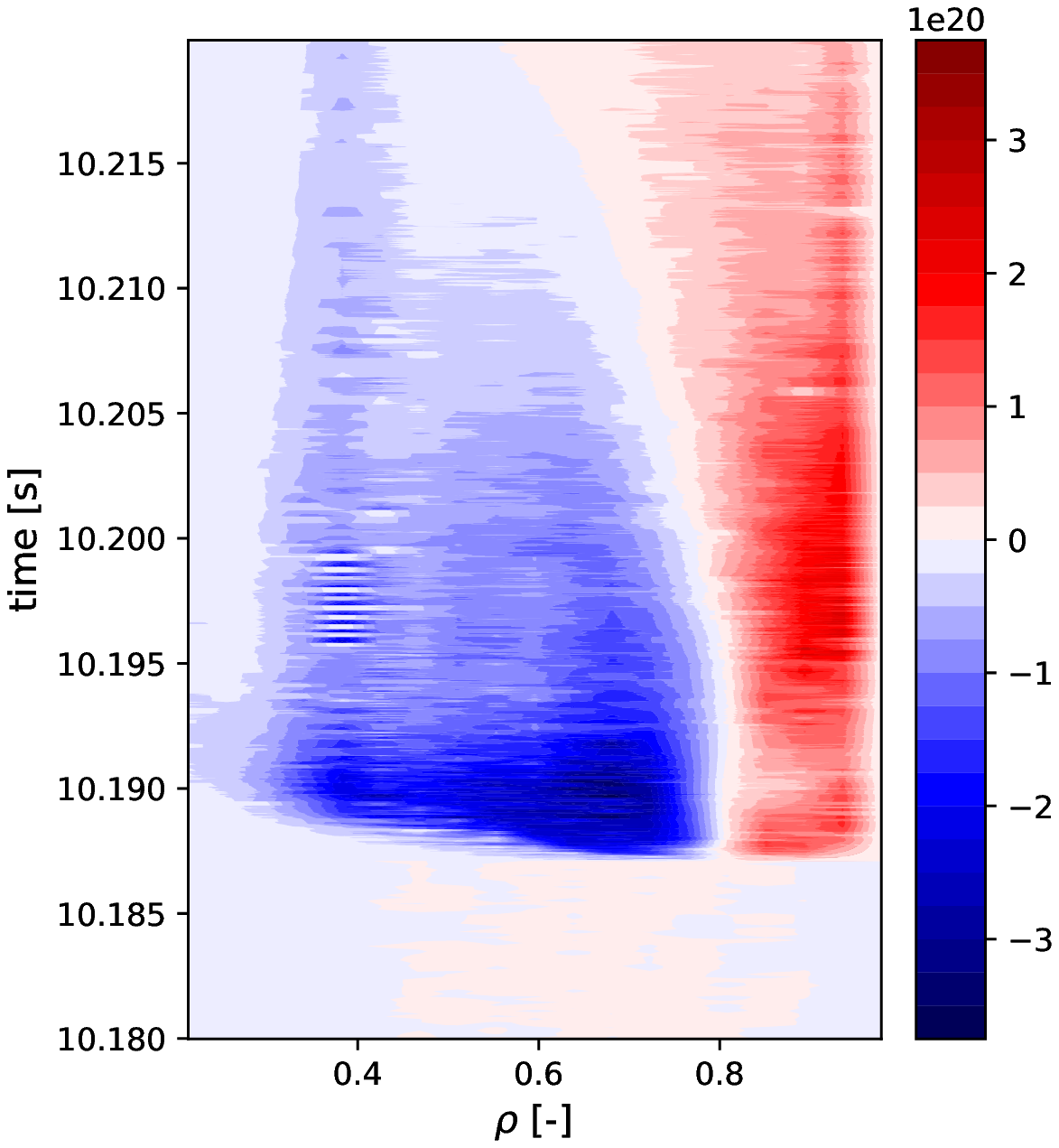}}

  \caption{Particle fluxes as function of time and $ \rho $ as predicted by QuaLiKiz, expressed as number of particles $ [m^{-2} s^{-1}] $. a) shows the electron particle flux, b) the hydrogen particle flux and c) the deuterium particle flux. Neglecting a small contribution from the impurities, the sum of the three plots gives zero. Warmer colors represent more outwards directed fluxes. A small amount of values are larger (or smaller) than the selected limits for the colorbar. Those value are saturated with the warmest (coldest) possible color.}
  \label{fig:fluxes_plots}
\end{figure*}

In the experiment the deuterium is injected at cryogenic temperature and following pellet ablation and ionization, is then heated by collisions with the hydrogen. Immediately after the pellet injection, the plasma is therefore a mix of hot hydrogen and cold deuterium. In the model hydrogen and deuterium are instead supposed to instantly thermalize to the average temperature between the two isotopes. The neutron rate is strongly dependent on the deuterium temperature, so a delay between the pellet injection and the rise in the neutron rate is to be expected in the experiment, but not in the model.

Proper modelling of this effect would require the currently non available option of having different temperatures for different ions in JINTRAC. Exact calculations are therefore left for future work. Utilizing simple energy considerations, it is still possible to infer the order of magnitude of the expected temporal shift. Taking the total ion heating as heating power, the time required to heat all the deuterium particles in the pellet to the average plasma temperature can be calculated. This exercise results in a $t_{shift} \sim 30ms $. Such value is consistent with the observed time shift between the modelled and the experimental neutron rate.

All the assumptions made in the integrated modelling can impact both the absolute value and the temporal evolution timescales of the neutron rate. Since the absolute value is ultimately the result of a fit, for each assumption the important sensitivity is on the time evolution. This is crucial. Extensive tests are carried out, finding in general a small impact on the timescales of the deuterium penetration. The impact of two of the pedestal parameters is shown in Figure \ref{fig:nDevolution}. Other sensitivities included equilibrium, impurities, radiation, critical pedestal pressure and ad hoc electromagnetic stabilization. 

The deuterium transport timescale is comparable to the energy confinement time. In particular, the rapid evolution of the neutron rate after the first pellet is correctly reproduced in the model. This timescale depends on the turbulent regime and the agreement is a validation of the fast deuterium penetration and of both QuaLiKiz and HPI2. The resilience of the fast time scale suggests a high reliability of the isotope penetration predictions in this scenario.

The results are a consequence of the turbulence regime identified by QuaLiKiz. Depending on the radial position and on the phase of the pellet cycle, different modes are excited. TEM is found by QuaLiKiz to be the dominant instability following pellet injection outside $ \rho = 0.8 $, where most of the pellet is ablated. The large negative density gradient causes a large particle flux directed outwards, in line with expectations from previous works \cite{Angioni2017}. However, in spite of this strong outward flux, pellet fuelling as observed by the inward deuterium penetration is achieved.

The particle fluxes as predicted by QuaLiKiz are presented in figure \ref{fig:fluxes_plots}. Outside $ \rho = 0.8 $, where TEM is the dominant instability, all the fluxes are outwards and the electron flux is the sum of the hydrogen and deuterium fluxes. The two ion particle fluxes are roughly proportional to their relative concentrations. For $ \rho < 0.8 $, where ITG is destabilized, the deuterium flux is the largest and is directed inwards. Ambipolarity is maintained by a smaller inward electron particle flux and a outwards hydrogen flux. Note that the expectation is not $ \Gamma_D/\Gamma_e \sim 10 $ at all times and across the entire profile. Pinch and diffusion vary in a non trivial way during the pellet cycle and can sometimes partially balance each other. For example, just after the pellet and between $ 0.6 < \rho < 0.8 $, where the electron density gradient is very large, $ D_e $ can be approximately a factor two larger. Still, $ D_i, V_i > D_e, V_e $ holds everywhere and $ D_i, V_i \gg D_e, V_e $ almost everywhere. As is clear from Figure \ref{fig:fluxes_plots}, the ion particle fluxes are significantly larger and the mixing is indeed observed. The key point is that the pellet perturbation initiates a transient transport event, where the large ion transport coefficients in the ITG regime enables the short isotope mixing timescale.

In future reactors the collisionality will be lower and the heating will be dominated by electron heating from fusion-generated alpha particles. The turbulence regime is predicted to be mixed ITG-TEM \cite{fable2019}. It is therefore important to model such a regime to assess the extrapolability of the fast isotope mixing effect to reactor-relevant plasmas. Some insight was gained here by repeating the same integrated modelling simulations while artificially reducing the collisionality input into QuaLiKiz towards reactor-relevant values. The detailed results are not shown for brevity. The turbulent regime is modified to a mixed ITG-TEM regime and the density peaking increases. However, ITG is still destabilised by the pellet at low wavenumbers and significantly contributes to the ion heat and particle transport. The timescale for the deuterium penetration is almost unchanged, confirming that it only depends on ITG being sufficiently destabilized and not on it being the sole dominant instability. The same qualitative result is obtained by changing the size and frequency of the pellets, with little or no impact on the isotope mixing timescale.

\begin{figure}
	\centering
	\includegraphics[width=1.0\linewidth]{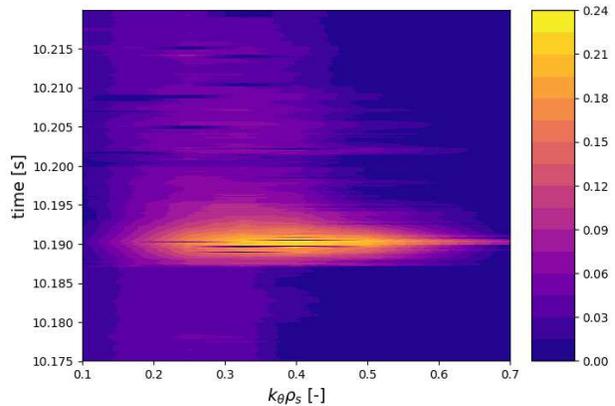}
	\caption{Growth rates in GyroBohm units for $ \rho = 0.68 $ during the first pellet cycle. $ k_{\theta} \rho_{s} $ is the normalized wavenumber $ k_{\theta} \frac{\sqrt{T_{e}m_{i}}}{q_{e}B} $, with $ m_{i} $ being hydrogen mass.}
	\label{fig:turbulence_visualization}
\end{figure}

\section{Gyrokinetic analysis} \label{GENE}

\subsection{Linear gyrokinetic analysis} \label{Linear}

\begin{figure}

  \subfloat[]{\includegraphics[width=.99\linewidth]{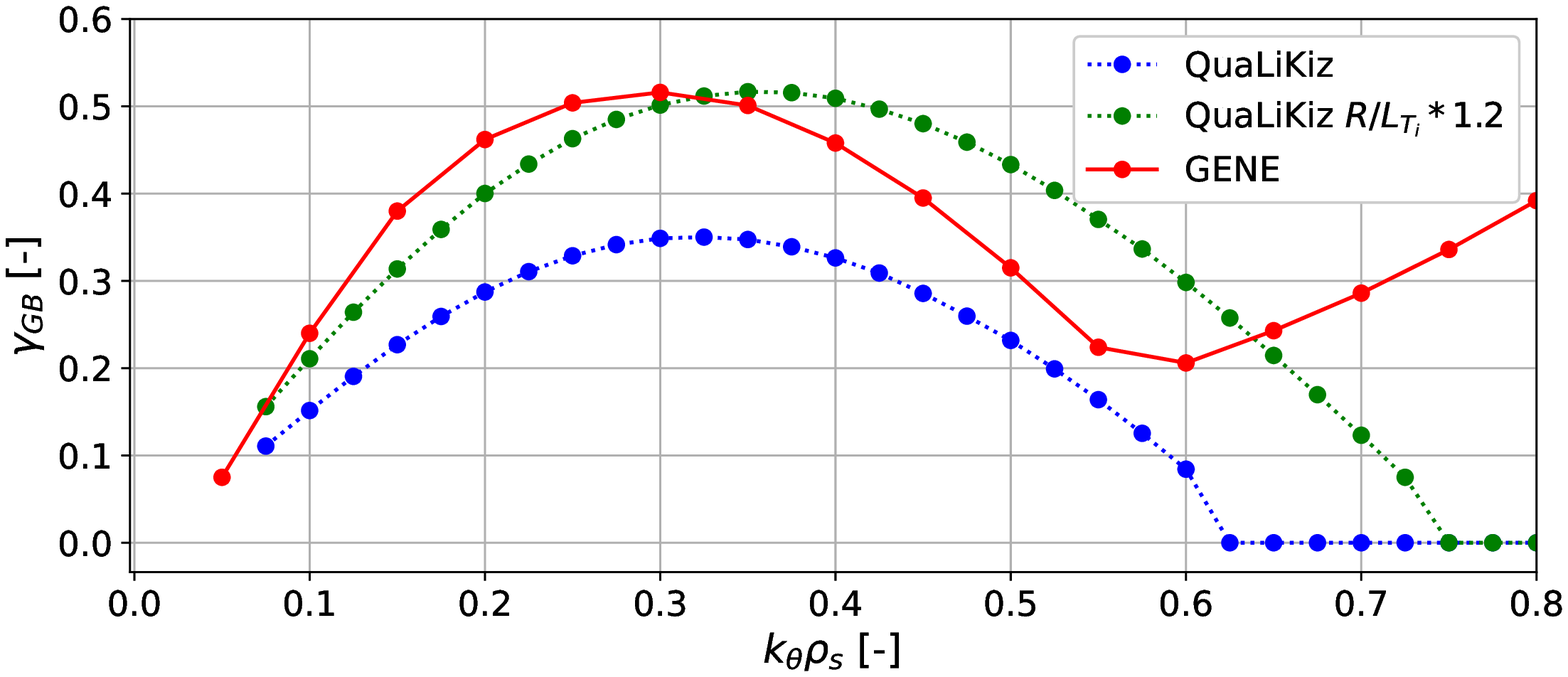}}
  \hfill
  \subfloat[]{\includegraphics[width=.99\linewidth]{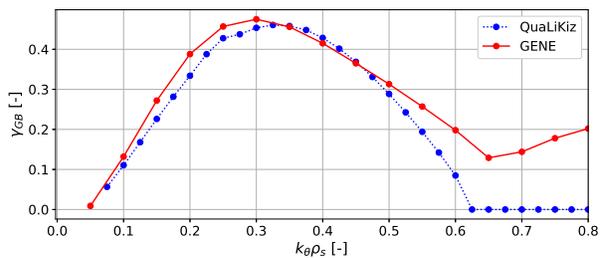}}

  \caption{Comparison between the normalized growth rates from GENE (red solid line) and QuaLiKiz (green and blue dotted lines). The parameters for both scans are taken from the JINTRAC simulation at $ \rho = 0.7 $ just before (a) and 10ms after (b) the first pellet. In the upper panel, the blue and green points indicate a simulation with nominal and $ 20\% $ increased $ R/L_{T_{i}} $ respectively. In GENE, the mode switches from the ion to the electron diamagnetic direction for $ k_{\theta}\rho_{s} > 0.6 $ in the upper panel (a) and for $ k_{\theta}\rho_{s} > 0.5 $ in the lower panel (b). $ `s` $ indicates the main specie, hydrogen in this case. The modes for QuaLiKiz are in the ion diamagnetic direction over the whole spectrum.}
	\label{fig:GENEvsQLK}
\end{figure}

The temporal behaviour of the instabilities, as predicted by QuaLiKiz, with ITG destabilized over a broad spectrum just after the pellet, is illustrated in figure \ref{fig:turbulence_visualization}. Note that the pellet is injected at $ t = 10.187s $ In this case, immediately after the pellet injection, the cooling caused by the adiabatic ablation of the pellets results in a locally steeper $ R/L_{T} $ gradient for $ \rho < 0.8 $. This balances the stabilizing impact of negative $R/L_n$ which occurs for ITG modes with kinetic electrons. This is key since the fast mixing of the deuterium depends on the ITG drive. To verify this important observation, eigenvalue solutions from QuaLiKiz is compared with linear calculations using the higher fidelity code GENE at $ \rho = 0.7 $. The growth rate comparison is shown in figure \ref{fig:GENEvsQLK} for time slices just before and 10ms after the pellet. The input parameters are taken directly from the integrated modelling simulation and are reported in Table \ref{tab:StandaloneParameters}. For simplicity, impurities and rotation are not included in this gyrokinetic comparison. The GENE settings are chosen to match the QuaLiKiz assumptions as close as possible. $ \beta $ is set to zero in GENE and the $ s-\alpha $ geometry model is used. The ad-hoc electromagnetic stabilization in QuaLiKiz is simply a modification of the input ion temperature gradient, so running both codes with the same $ R/L_{T_i} $ is equivalent to not including electromagnetic effects in QuaLiKiz. There is a difference in how collisions are treated, since GENE employs a linearized Landau-Boltzmann operator, and QuaLiKiz a Krook-like operator for trapped electrons.

\begin{table*}
    \caption{Simulation parameters used in the linear and nonlinear analysis, corresponding to the integrated modelling parameters $ 10 ms $ after the first pellet. $ x $ is defined as $ x = r/a $ , where $ r $ is minor radius of the flux surface, and $ a $ is the minor radius of the last-closed-flux-surface. q is the safety factor and s the magnetic shear. Electron and ion temperatures are given in eV.}
    \centering
    \smallskip
     \begin{tabular}{c|c|c|c|c|c|c|c|c|c|c|c}
	    $ \rho $ & x & q & s & $ n_e $ & $ T_e $ [eV] & $ T_i $ [eV] & $ R/L_{T_{e}} $ & $ R/L_{T_{i}} $ & $ R/L_{n_{e}} $ & $ R/L_{n_{H}} $ & $ R/L_{n_{D}} $ \\
        \hline
	    0.7 & 0.75 & 2.32 & 1.49 & 4.21 $ 10^{19} $ & 738 & 814 & 14.9 & 12.9 & -7.96 & 0.93 & -18.0 \\

	\end{tabular}
	\label{tab:StandaloneParameters}
\end{table*}

The growthrates increase by a factor of 3 immediately after the first pellet, as visible in figure \ref{fig:turbulence_visualization}. The increase is instead more moderate in figure \ref{fig:GENEvsQLK}, comparing the two QuaLiKiz simulations. This difference is due to the smoothing applied when extracting the values from the JINTRAC simulation, since the output is generated every $ \sim 5ms $.

The sudden change in the gradients caused by the pellet moves the system far from the threshold, resulting in large growthrates. This causes large fluxes, which quickly flatten the most extreme gradients. The process lasts for $ \sim 5ms $. The relatively slower evolution that follows, with the density profile going from hollow to peaked, shows a more moderate increase in the growth rates and fluxes closer to the stationary state values. The parameters chosen for the GENE simulation are representative of this phase.

In the pre-pellet phase, ITG modes dominate for $k_\theta\rho_s<0.6$. QuaLiKiz predicts lower growth rates than GENE, but with a very similar spectral shape. An increase of the ion temperature gradient by $ 20\% $ is sufficient for QuaLiKiz to retrieve the GENE growth rates. TEM is found to be unstable by GENE and stable by QuaLiKiz for $ k_{\theta}\rho_{s} > 0.6 $. This is most likely due to the collisional operator in QuaLiKiz, which tends to over-stabilize TEM and is currently being upgraded. Furthermore, TEM is responsible for only a small fraction of the total transport in this case, and the presence of TEM does not affect the central result of the fast isotope mixing, since in a mixed ITG - TEM regime both ion and electron particle transport are expected to be fast \cite{Marin2020}. In the post-pellet phase, ITG again dominates in the transport driving region $k_\theta\rho_s<0.6$. QuaLiKiz and GENE growth rates agree very well at nominal input parameters in this region. TEM is the dominant mode in GENE for $k_\theta\rho_s>0.6$. The key result is that indeed ITG is destabilized in GENE in presence of a positive density gradient in the post-pellet phase, validating the QuaLiKiz predictions.

\begin{table}
    \caption{Details on the grid used for the nonlinear GENE simulation. $ n_x $, $ n_y $, $ n_z $, $ n_v $ and $ n_w $ represent the number of grid points respectively for the radial, bi-normal, parallel, $ v_{||} $ and magnetic moment dimensions. $ k_{y,min} $ is the minimum value $ k_y $ mode in the simulation, normalized to the inverse gyroradius. $ L_x $, $ L_v $ and $ L_w $ are the extension of the simulation box in the radial, $ v_{||} $ and magnetic moment directions, normalized to the inverse gyroradius, thermal velocity and $ T_s/B $ respectively}
    \centering
    \smallskip
     \begin{tabular}{c|c|c|c|c|c|c|c|c}
	    $ n_x$  & $ n_y $ & $ n_z $ & $ n_v $ & $ n_w $ & $ k_{y,min} $ & $ L_x $ & $ L_v $ & $ L_w $ \\
        \hline
	    128 & 32 & 32 & 64 & 15 & 0.05 & 100.0 & 3.0 & 9.0 \\

	\end{tabular}
	\label{tab:GENEgridnonlinear}
\end{table}

\begin{figure*}

  \subfloat[]{\includegraphics[width=.5\linewidth]{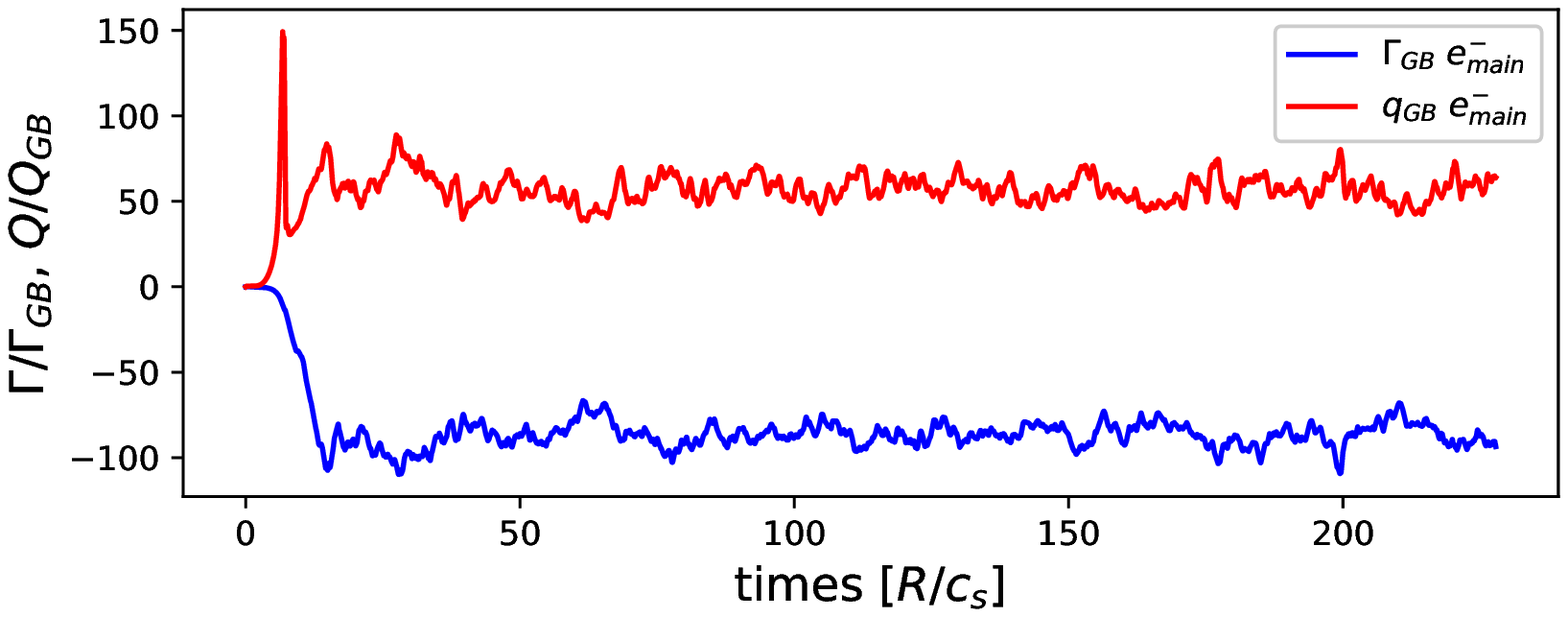}}
  \subfloat[]{\includegraphics[width=.5\linewidth]{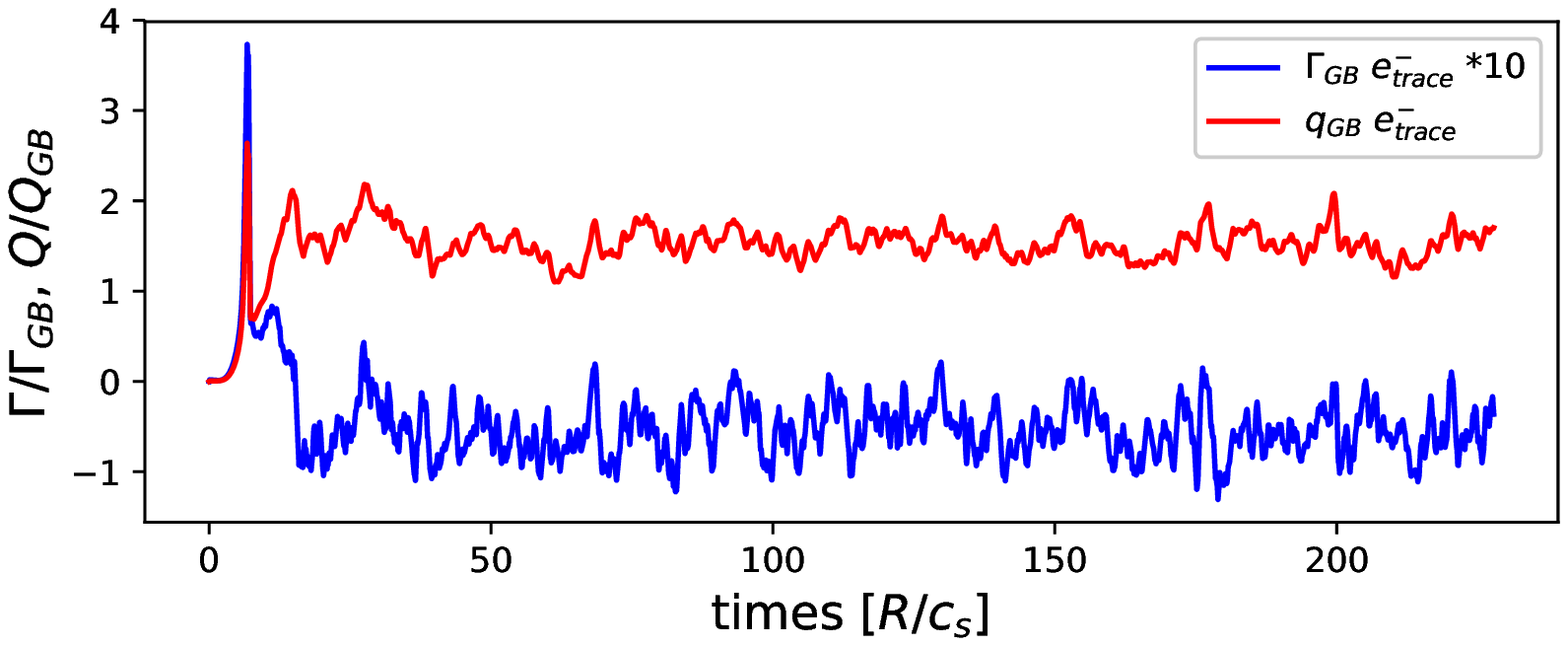}}
  \hfill
  \subfloat[]{\includegraphics[width=.5\linewidth]{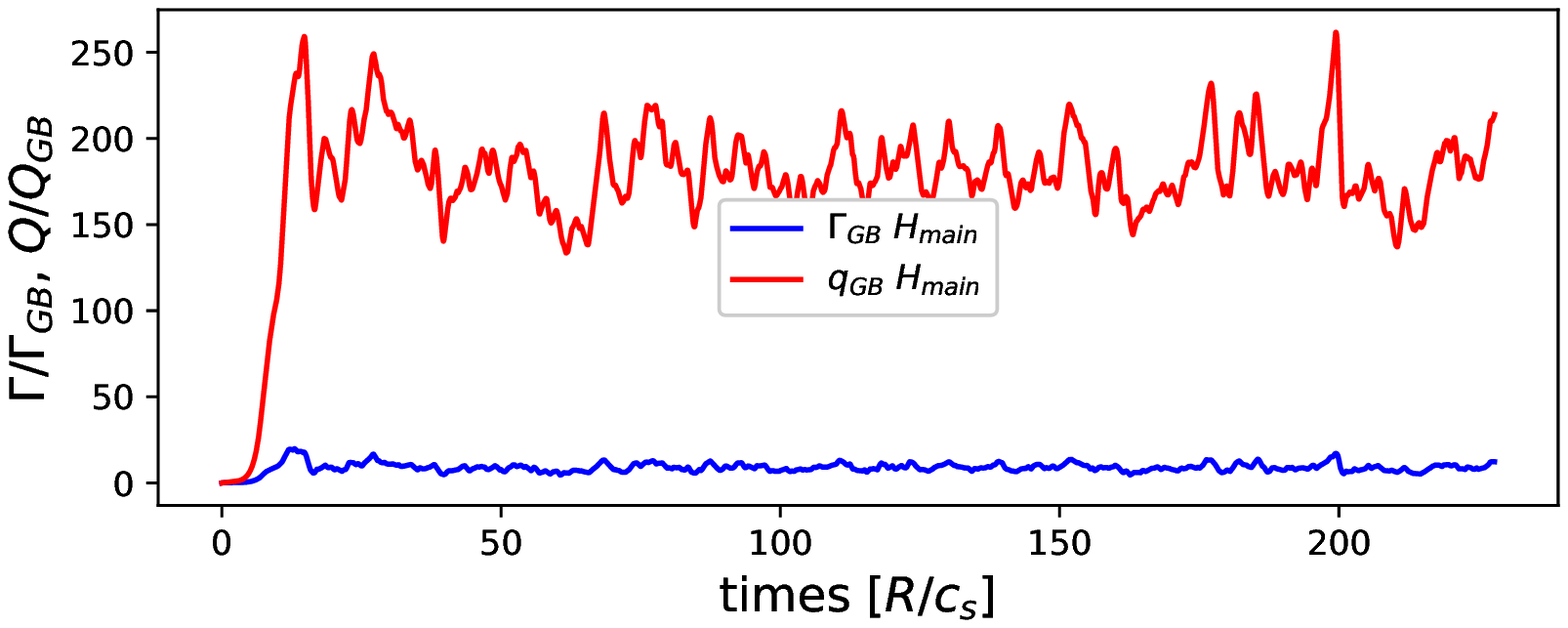}}
  \subfloat[]{\includegraphics[width=.5\linewidth]{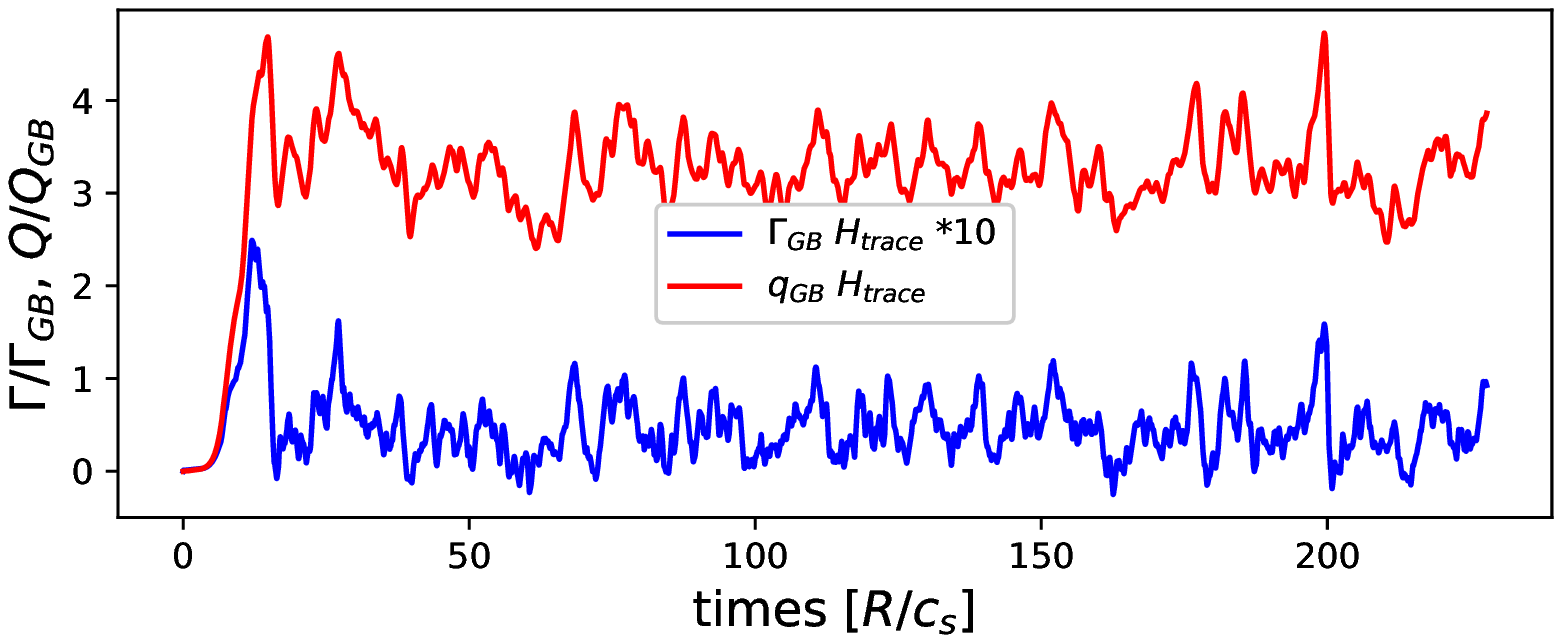}}
  \hfill
  \subfloat[]{\includegraphics[width=.5\linewidth]{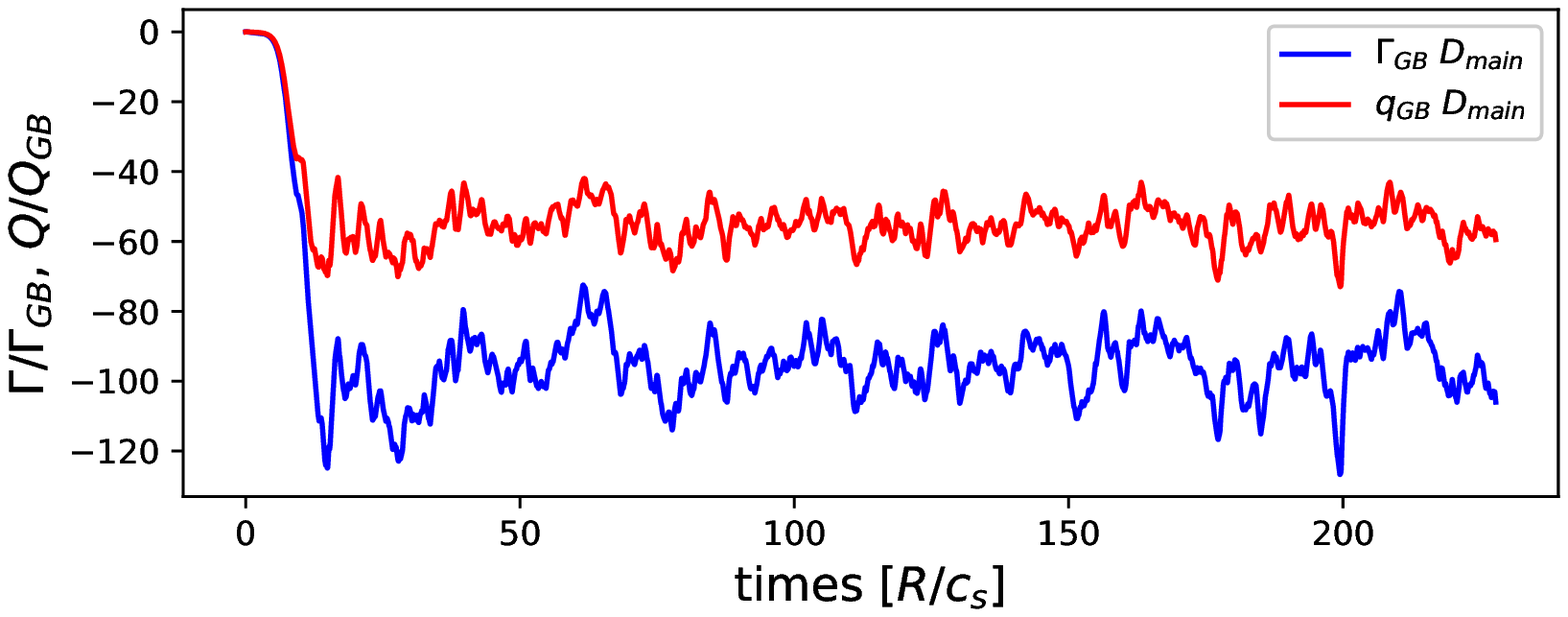}}
  \subfloat[]{\includegraphics[width=.5\linewidth]{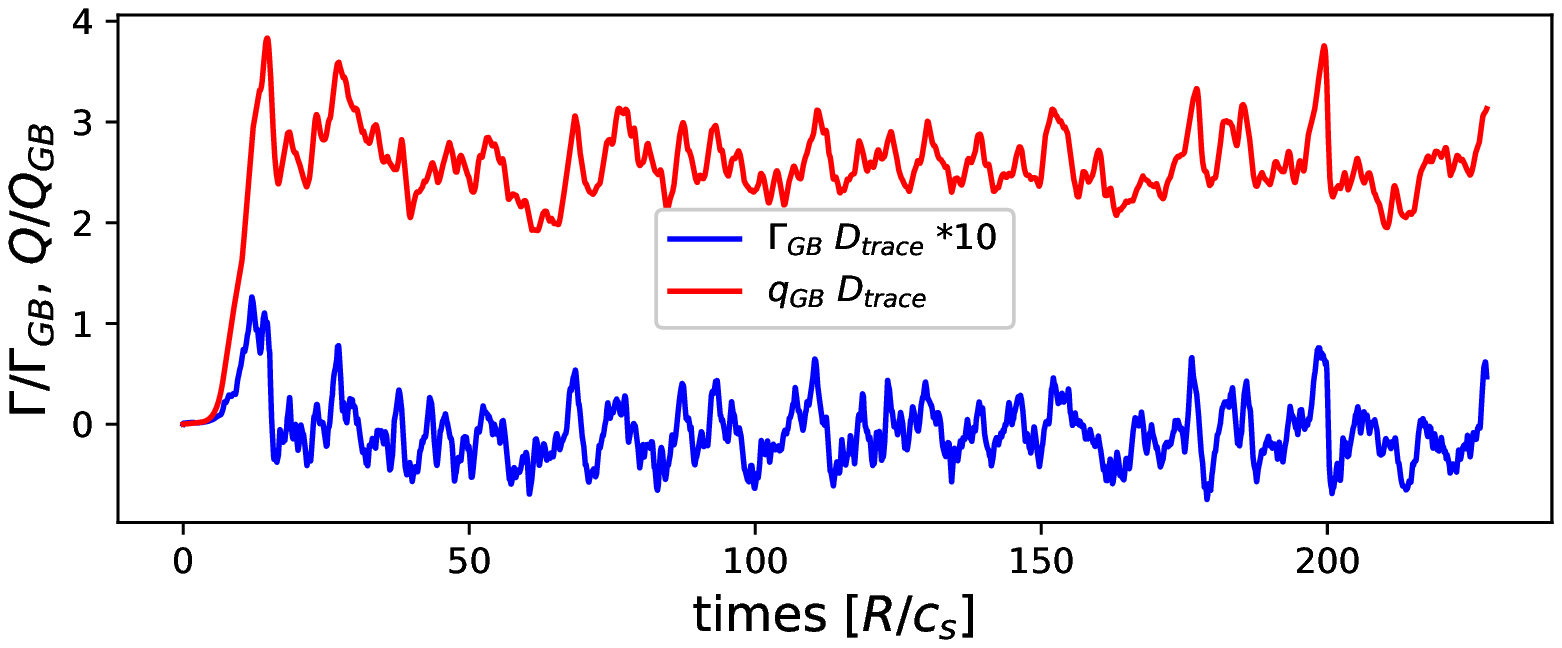}}
  \hfill
  
  \caption{Heat and particles nonlinear fluxes as predicted by GENE. The main ions are shown on the left hand side, while the traces are on the right hand side. From top to bottom, in order, the electron, hydrogen and deuterium species. The particle fluxes are normalized to $ \Gamma_{GB} = c_{ref} n_{e} (\rho^*_{ref})^2 $. $ c_{ref} = \sqrt{T_e/m_p} $ is the reference velocity, $ n_e $ the electron density and $ \rho^*_{ref} = (c_{ref}/\Omega_{ref})/R $ the normalized gyro-radius. $ \Omega_{ref} = (q_e B)/{m_p c} $ is the gyro-frequency, with $ q_e $ the electron charge, $ m_p $ the proton mass, c the speed of light and R the major radius. The heat fluxes are normalized to $ Q_{GB} = c_{ref} p_{e} (\rho^*_{ref})^2 $, with $ p_{ref} $ the reference pressure. The times are normalized to $ R/c_{ref} $.}
  \label{fig:nonlinear_fluxes}
\end{figure*}

\subsection{Nonlinear gyrokinetic analysis} \label{Nonlinear}

Further validation of the QuaLiKiz predictions is explored through full nonlinear GENE simulations. Since the primary interest is on the particle transport, ion-scale simulations are deemed to be sufficient. For simplicity, the same assumptions as in the linear calculations are made: $ s -\alpha $ geometry, $ \beta = 0 $, no impurities and no rotation. The nominal input parameters for the nonlinear simulation are the same as set in the linear simulation. Grid resolution details are found in Table \ref{tab:GENEgridnonlinear}. Extensive numerical convergence tests were carried out for $n_x$, $n_y$, $k_{y,min}$, $n_v$ and $n_w$ inputs, confirming that the grid chosen is indeed sufficient to resolve the physics under consideration. The evolution of the heat and particle fluxes for the various species is shown in figure \ref{fig:nonlinear_fluxes}.

The transport coefficients $ D_s, V_s $ are calculated by adding extra electron, deuterium and hydrogen trace species. The density of each trace is set to $ 1\% $, while the density gradient is set to zero. $ D_s, V_s $ are then calculated using

\begin{equation}
V_s = \Gamma_{trace}/n_{trace},
\end{equation}

\begin{equation}
D_s =  (\Gamma_{main} - n_{main} V_s) L_{n,s}/n_{main}.
\end{equation}

$ \Gamma_{main} $ is the particle flux for $ e^- $, H and D, while $ \Gamma_{trace} $ is the particle flux for the respective trace. The same notation holds for $ n $, $ R_0 $ is the major radius and $ R/L_{n,s} $ the normalized logarithmic density gradient. The saturated fluxes are obtained by averaging from $ t = 40 $ to the end of the simulation. Fluxes and transport coefficients are summarized in Table \ref{tab:GENEfluxesnonlinear}.

Agreement between the GENE nonlinear simulation and power balance from the integrated modelling run, is achieved by increasing $ R/L_{T_i} $ by $ 20\% $. This corresponds to $\sim30\%$ lower heat fluxes than in QuaLiKiz, since in integrated modelling the QuaLiKiz run also included the stabilizing effect of rotation and impurities. This agreement in heat fluxes and heat flux ratios between QuaLiKiz, nonlinear-GENE, and experimental power balance, in this positive density gradient regime, is not trivial and can be considered a highly encouraging validation in itself. Consistently with the linear results, frequency analysis in the nonlinear run (and not shown for brevity) shows that the turbulence is predominantly ITG with subdominant TEM.

With respect to the particle transport, some quantitative differences are observed between nonlinear-GENE and QuaLiKiz, although the result of fast deuterium penetration is still obtained in the nonlinear simulation. In comparison to QuaLiKiz, the electron flux is significantly more inward. This is due to an increased electron particle diffusion term in GENE compared to QuaLiKiz. This may arise due to the increased TEM drive in GENE, which is expected to increase electron particle transport coefficients \cite{Bourdelle2018a}. The increased electron inflow is compensated by a reduced hydrogen outflow in GENE compared to QuaLiKiz. However, crucially, the deuterium inflow is still large in the GENE simulation, even larger than in QuaLiKiz. This maintains the key result of fast deuterium penetration in the post-pellet phase. Furthermore, the GENE and QuaLiKiz ion diffusion coefficients are very comparable.

Considering the convective coefficients, a large difference is found for the ions, with $ V_i (GENE) \ll V_i (QuaLiKiz) $, while the inward pinch for the electrons is slightly larger in GENE. This disagreement is not completely unexpected since differences in particle transport were previously observed between quasilinear and nonlinear simulations in this positive density gradient regime \cite{Baiocchi2015}. Further investigation on the origin of the discrepancy will be object of future work.

\begin{table*}
    \caption{Comparison between the fluxes and the transport coefficients as predicted by nonlinear GENE and QuaLiKiz. All the values are reported in SI units. q is the heat flux and e, H and D stand for electrons, hydrogen and deuterium. The parameters for the simulations have been taken from the integrated modelling, $ 10 ms $ after the first pellet. The $ s-\alpha $ geometry model is used, $ \beta $ is set to zero in GENE and impurities and rotation are not included.}
    \centering
    \smallskip
    \begin{tabular}{c|c|c|c|c|c|c|c|c|}
	    code & \multicolumn{2}{|c}{$ \Gamma $ $ [m^{-2} s^{-1}] $} & \multicolumn{2}{|c}{$ q $ $ [W m^{-2}] $} & \multicolumn{2}{|c}{$ D $ $ [m^{-2} s^{-1}] $} & \multicolumn{2}{|c}{$ V $ $ [m s^{-1}] $} \\ 
	    \hline
	     & $ \Gamma_e $ & $ -2.57 $ $ 10^{20} $ & $ q_e $ & $20.0 $ $ 10^3 $ & $ D_e $ & 2.18 & $ V_e $ & -0.39 \\
	    GENE & $ \Gamma_H $ & $ 2.65 $ $ 10^{19} $ & $ q_H $ & $ 63.6 $ $ 10^3 $ & $ D_H $ & 2.93 & $ V_H $ & 0.31 \\
         & $ \Gamma_D $ & $ -2.84 $ $ 10^{20} $ & $ q_D $ & $ -19.5 $ $ 10^3 $ & $ D_D $ & 2.41 & $ V_D $ & -0.08 \\
        
        \hline
        
         & $ \Gamma_e $ & $ -0.775 $ $ 10^{20} $ & $ q_e $ & $ 31.4 $ $ 10^3 $& $ D_e $ & 1.03 & $ V_e $ & -0.19 \\
	    QuaLiKiz & $ \Gamma_H $ & $ 1.27 $ $ 10^{20} $ & $ q_H $ & $ 97.2 $ $ 10^3 $ & $ D_H $ & 3.39 & $ V_H $ & 4.79 \\
         & $ \Gamma_D $ & $ -2.046 $ $ 10^{20} $ & $ q_D $ & $ -25.8 $ $ 10^3 $ & $ D_D $ & 2.61 & $ V_D $ & 5.16 \\
        
	\end{tabular}
	\label{tab:GENEfluxesnonlinear}
\end{table*}

\section{Conclusions} \label{Conclusions}

The JINTRAC integrated modelling framework with the turbulent transport model QuaLiKiz as the turbulent transport model and HPI2 as the pellet deposition model successfully reproduced observations over multiple pellet cycles in JET mixed-isotope experiments. Good agreement on the density profile evolution and on the neutron rate timescales was achieved. The compensation between $ R/L_{n} $ stabilization and $ R/L_{T} $ destabilization was shown to lead to maintained ITG drive and allow prompt deuterium penetration on energy confinement timescales following each pellet injection throughout the pellet train. The key QuaLiKiz prediction of ITG instability in post-pellet negative $R/L_n$ regimes was verified by linear and nonlinear GENE simulations. The same core modelling approach presented in this Paper can be used to predict the timescale for the fuel penetration in ITER and future reactors, with optimistic preliminary results with regard to fuelling capability and burn control.

\section{Acknowledgements} \label{Acknowledgements}

This work has been carried out within the framework of the EUROfusion Consortium and has received funding from the Euratom research and training programme 2014-2018 and 2019-2020 under grant agreement No 633053 and from the RCUK [grant number EP/P012450/1]. The views and opinions expressed herein do not necessarily reflect those of the European Commission.

\printbibliography

\end{document}